# RADIO FREQUENCY INTERFERENCE MITIGATION AT THE WESTERBORK SYNTHESIS RADIO TELESCOPE: ALGORITHMS, TEST OBSERVATIONS, AND SYSTEM IMPLEMENTATION

W. A. Baan, P. A. Fridman, and R. P. Millenaar
ASTRON, Postbus 2, NL-7990AA, Dwingeloo, Netherlands; baan@astron.nl, fridman@astron.nl, millenaar@astron.nl

## ABSTRACT

The sensitivity of radio astronomical stations is often limited by man-made radio frequency interference (RFI) due to a variety of terrestrial activities. An RFI mitigation subsystem (RFIMS) based on real-time digital signal-processing is proposed here for the Westerbork Synthesis Radio Telescope based on a powerful field programmable gate array processor. In this system the radio astronomy signals polluted by RFI are "cleaned" with the RFIMS before routine back-end correlation processing takes place. The high temporal and frequency resolution of RFIMS allows the detection and excision of RFI better than do standard radio telescope back-end configurations.

## 1. INTRODUCTION

The sensitivity and spectral performance of modern telescopes is increasingly affected by the worsening electromagnetic environment resulting from population density and increased spectrum use. At the same time, technology standards at modern radio telescopes have been raised significantly, and the noise of the receivers has been lowered. However, as a result of the presence of radio frequency interference (RFI) in the data, these potentials of sensitivity cannot be reached at all times, especially at frequencies below 1 GHz. The presence of detrimental RFI in the data translates into effective data loss and lower observing efficiency.

Several methods of RFI mitigation in radio astronomy have been proposed and implemented in recent years. In practice, there is no universal RFI "bullet-proof" mitigation method and successful mitigation can only be achieved using a combination of various engineering countermeasures. Knowledge of the local RFI environment and the use of an RFI database compiled at a particular site are valuable ingredients in combatting RFI signals. Single-dish observatories have applied a number of different methods to eliminate the effects of RFI, including: (1) scheduling observations during the night hours and periods of low RFI; (2) stopping the correlation process during periodic radiation bursts, such as those of radar systems; (3) front-end filtering; and (4) excising data during postprocessing that contain interfering signals. Of course, all these methods constitute effective data loss and reduced observing efficiency. In addition, the data obtained with interferometers are subject to decorrelation of RFI signals due to fringe rotation caused by the separation between antennas. Larger baselines give more robust suppression of nonastronomical signals of terrestrial origin.

Advances in digital techniques provide vast opportunities for RFI mitigation, and some observatories have experimented with mitigation systems (Barnbaum & Bradley 1998; Briggs et al. 2000; Ellingson et al. 2000; Mitchell & Bower 2001). A demonstrator RFI mitigation system has been developed at the Westerbork Synthesis Radio Telescope (WSRT) based on a Signatec PMP-8 digital signal processor (DSP) board and four-channel 12-bit high-speed (up to 50 Msamples s$^{-1}$) analog-digital and digital-analog converters (ADCs and DACs, respectively) (Fridman & Baan 2001; Fridman 2000). This system was used to operationally test several RFI mitigation algorithms. The principal methods used in these tests were (1) time-frequency analysis and outliers excision using thresholding, (2) RFI suppression using a "RFI estimation→subtraction" method, (3) adaptive RFI cancellation using a reference channel, and (4) using higher order statistics for analyzing the data. These DSP processing efforts of the real astronomical signals resulted in considerable RFI suppression both for continuum and spectral observations.

As a follow-up of this DSP-based system, a second-generation real-time RFI Mitigation System (RFIMS) has been proposed for implementation within the existing WSRT back-end infrastructure. The *real-time* processing was chosen for RFIMS for the following reasons:

1. Very often RFI signals are unstable, rapidly changing, and very strong. They may behave like bursts in temporal or frequency domains. The RFI may produce considerable distortions of averaged data at the correlator output if not "intercepted" before processing in the correlator. RFI at the antennas of such a large and sparse interferometer as the WSRT is sometimes only partly coherent, and spatial processing algorithms are not always effective. Therefore, the *real-time* time-frequency analysis of the signals at each antenna is a more flexible tool than postcorrelation processing.

2. Fringe stopping and delay compensation are intrinsic radio interferometer procedures, which are necessary to keep the signals received at different interferometer sites coherent. The modulation and group delay caused by these operations produces a destructive impact on RFI. This effect depends on the geometry "baseline–sky radio source of interest–RFI source" (Thompson 1982). The attenuation of RFI may be considerable (see Appendix D), and postcorrelation RFI mitigation processing is less effective.

3. A radio interferometer such as WSRT is often used in a *tied-array* mode (VLBI and pulsar observations), meaning that it works as a "single-dish" radio telescope without a total power detector (the whole bandwidth baseband is processed).

The system is designed so that one of the eight 20 MHz wide subbands in each intermediate-frequency (IF) polarization channel of each of the 14 telescopes gets processed independently between the IF to baseband conversing stage and the correlation stage. The full system consists of 28 separate processors. State-of-the-art high-performance processors are being used for this real-time processing stage. It should be noted that this system is essentially a multipath processing system that could be effectively applied at single-dish telescopes as well. This paper reports on the first operational results obtained with this RFIMS system.

## 2. RFIMS: THE RFI MITIGATION SYSTEM

The implementation of an RFI mitigation subsystem within an existing interferometric signal-processing back end is a delicate technical problem. To allow implementation, the subsystem should (1) be adapted to the input-output specifications and the schematics of the back end, (2) be precisely synchronized with the correlation processing, and (3) keep the useful radio source signals undistorted and maintain correct absolute and constant time relations of the multichannel system. After the last frequency conversion stage in the intermediate-to-video-converter (IVC) system, the baseband signals of bandwidth up to 20 MHz are processed in the RFI mitigation subsystem before they are presented to the radio interferometric correlator. The processed signals should be presented to the correlator as if there are no additional processing stages involved.

Using the experience obtained with the DSP-based demonstrator system, a second-generation RFI mitigation subsystem has been conceived for the WSRT. The design criteria used for the RFIMS system are based on the following considerations:

1. For processing large quantities of data in a similar manner, a high degree of parallelism, extensive pipelining, dynamic reconfigurability and the reallocation of computational units, upgradability, and scalability are required.
2. High-performance processing cores (DSP, Pentium IV) provide a small benefit if the data transfer overhead causes their processing power to drop below unacceptable limits. Therefore, the processing power and data transfer speeds are equally important for system evaluations.
3. The signal-processing literature is full of creative solutions to real-world problems. Often these solutions are unavailable to a designer because they do not map well to software-programmable DSP architectures. Similarly, ASIC solutions are not an option for reasons of timescales, economy of scale, and flexibility. On the other hand, field programmable gate arrays (FPGAs) give immediate access to a diverse range of potential solutions for high-performance digital processing systems. These devices maintain the flexibility of software-based solutions, while providing levels of performance that match ASIC solutions.

A block-diagram of the RFIMS structural buildup is presented in Figure 1. Baseband signals are digitized, processed, and then transformed back to analogue form to be processed further in the correlator. The main processing unit consists of an ADC→FPGA→DAC configuration. Conversion back to analogue is necessary at WSRT because the correlator has analogue baseband inputs, and modification of this equipment is not an immediate option.

A 12-bit ADC with a maximum sampling frequency of 125 MHz is used to digitize the baseband signals at the output of the IF-to-video converter. The maximum bandwidth of the baseband signals is 20 MHz. The digitized data are supplied to the signal processing engine, which contains an ALTERA STRATIX programmable logic device EP1S80B956-6 for the RFI real-time mitigation processing. The processor has 79,040 logical elements and 7,427,520 RAM bits and works with clock frequencies up to 200 MHz. It also contains 176 embedded ($9 \times 9$) multipliers. The processed data are then converted back to the analogue form using a 14-bit DAC, with a maximum conversion frequency of 165 MHz. The RFIMS processing may be bypassed completely using a cross-point switch.

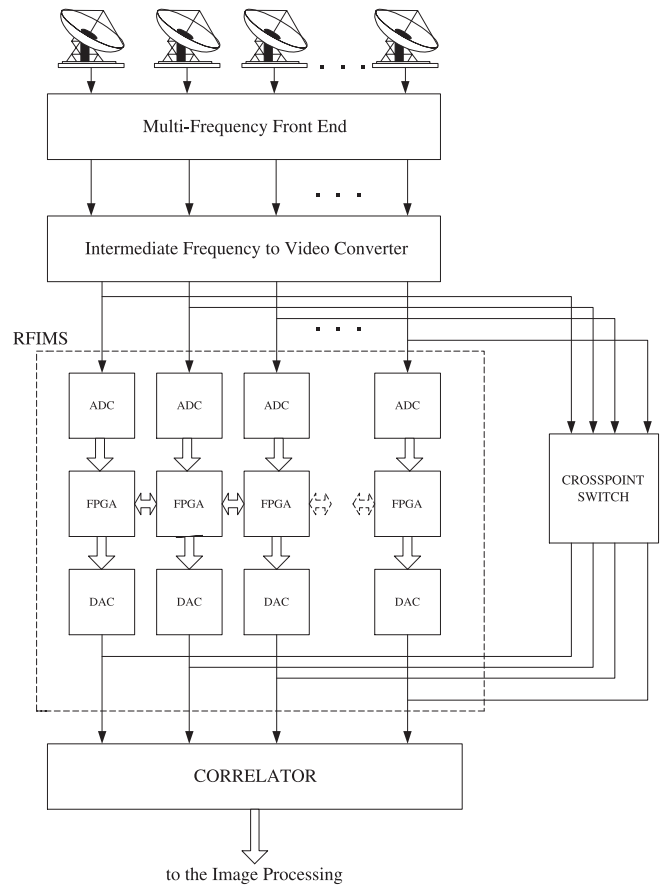

Fig. 1.—RFI mitigation subsystem at WSRT; FPGA-field programmable gate array.

During a particular observation, a configuration designed for a specific RFI mitigation algorithm is loaded in the FPGAs. The algorithm is adapted to and optimized for observation type and the particular RFI environment. If the situation changes or a new observational program is to be started, a new configuration can be loaded. The observatory staff will need to monitor the RFI environment at each of the antennas and evaluate the effectiveness of the particular RFI mitigation algorithms. At a later stage, a knowledge-based system may take over this human intervention.

All equipment is based on off-the-shelf VME modular components (COTS) as a mainframe. The processing modules are all connected with each other and can be set up to exchange data. This option is needed for applications based on spatial filtering algorithms and for synchronization of calculations in different parallel channels as in a *wave array processor*.

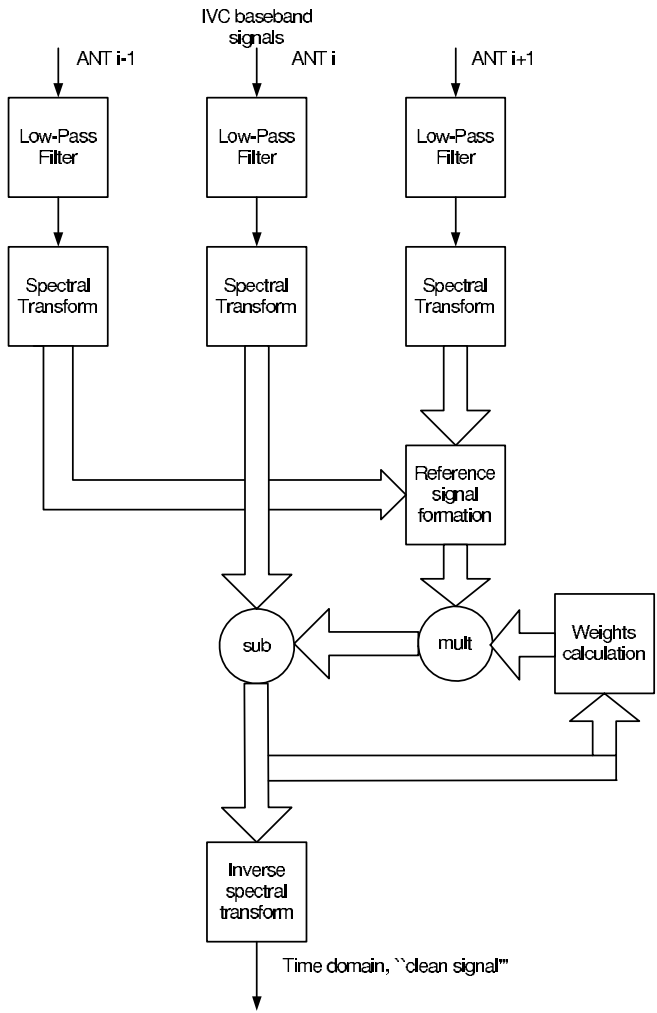

Fig. 2.—Block diagram of the adaptive RFI cancellation.

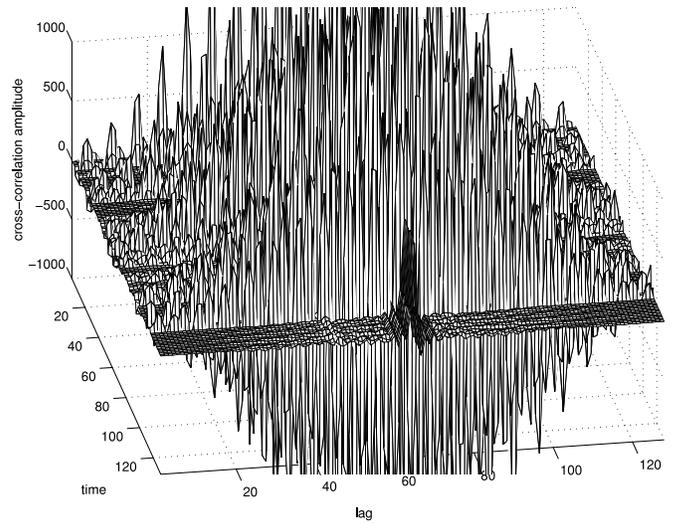

Fig. 3a

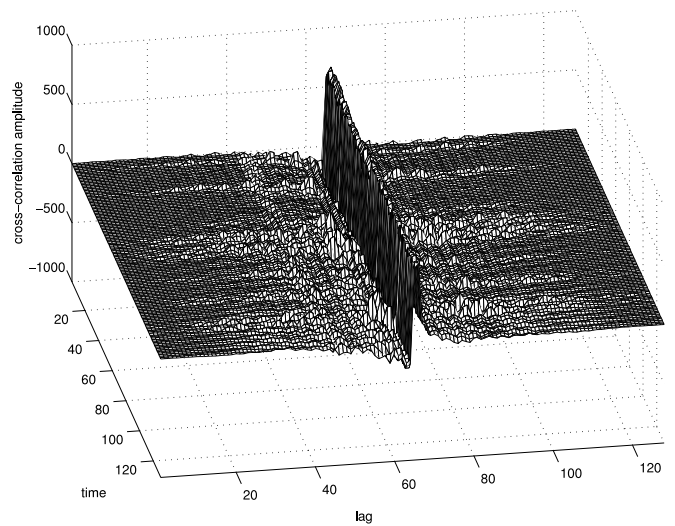

Fig. 3b

Fig. 3.—Observation at WSRT No. 10300911, 2003 January 29, source 3C48, frequency 337 MHz, bandwidth 10 MHz, DZB correlator, 60 s integration time for each of the 131 records ($\approx 2h$). RFI signals were suppressed at channels RT5X and RT7X, and not suppressed at channels RT4X and RT6X. The time-frequency presentations of the *cross-correlation amplitudes* are given: (*top*) 4X6X, (*bottom*) 5X7X. Note that both interferometers have equal baselines. Attenuation of this sporadic RFI in the channel 5X7X reaches $\approx 20 db$.

For WSRT the minimum number of the channels is 28 with 14 antennas and two polarizations. In the first instance, just one of the eight 20 MHz frequency channels will be processed before correlation. Eventually the system may be expanded to up to 224 channels to cover eight 20 MHz frequency bands from each antenna.

### 3. ALGORITHMS

Two main algorithms have been implemented in the FPGA as first trial processes.

*Time-frequency analysis of baseband signals and RFI excision*.—Thresholding has been used in both the temporal domain and the frequency domain. After a real-time fast Fourier transform (FFT) in the FPGA, the running power spectrum is calculated, followed by a threshold detection of the RFI in the time-frequency plane (see Appendices A and B). The detection of RFI *transients* of unknown form in the time-frequency plane can be made effectively with a cumulative sum (CUSUM) procedure, which maps well onto the FPGA structure. However, a thresholding procedure requires knowledge of the quiescent noise variance in each of the frequency bins, which is the spectral density of the system noise in the absence of RFI. This variance could be different for each baseband because of the variability of the transfer functions of the low-pass filters in the WSRT IVC system. A nonparametric procedure to achieve this is median filtering of the power spectrum to obtain a reference power spectrum, which determines the threshold level in the spectral domain.

*RFI suppression using adaptive noise cancelation (ANC) techniques*.—The 14 WSRT antennas form a linear, sparse but regular array with separations of 144 m. Because of space constraints, the installation of reference antennas and receivers near the focal region of the array elements is not feasible. Instead, a combination of spatial filtering, adaptive nulling, and ANC techniques is proposed for WSRT (Widrow & Stearns 1985; Haykin 1996); see the block diagram in Figure 2. The effectiveness of ANC is considered in Appendix C.

All processing is performed in the frequency domain. The signals from the IVC system, where array-based phase-rotation (the fringe-stopping routine) has been done, are then

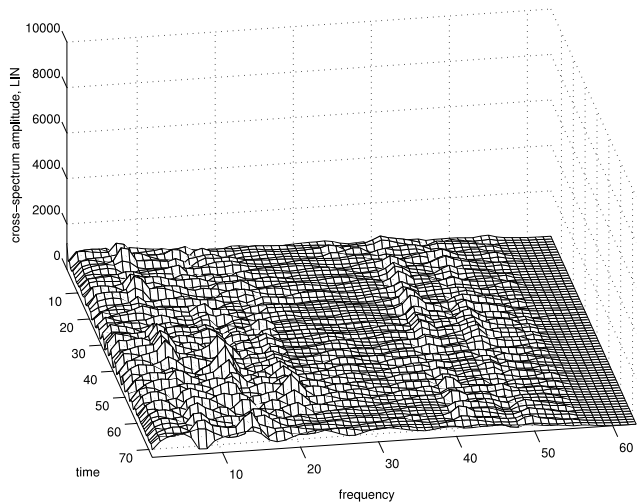

Fig. 4a

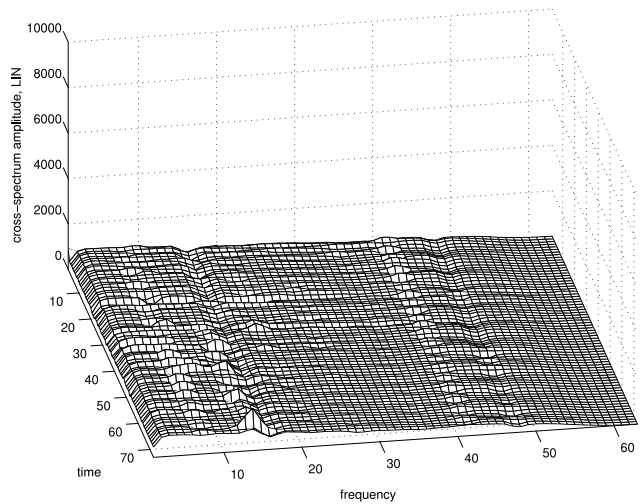

Fig. 4b

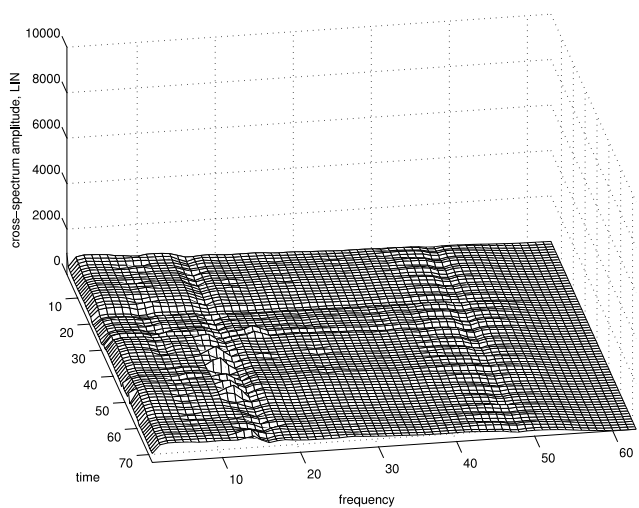

Fig. 4c

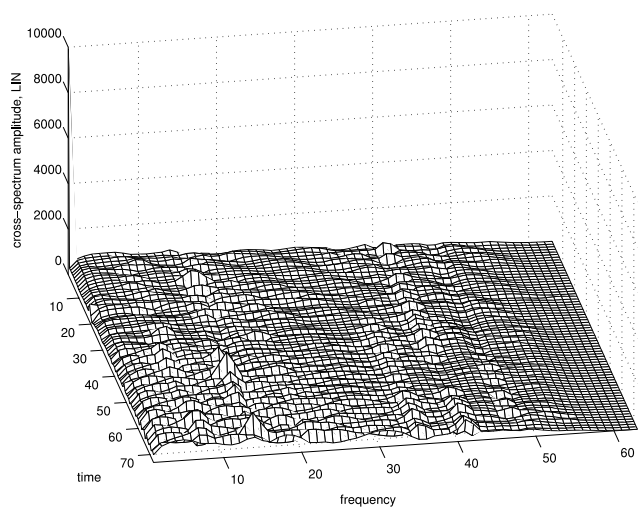

Fig. 4d

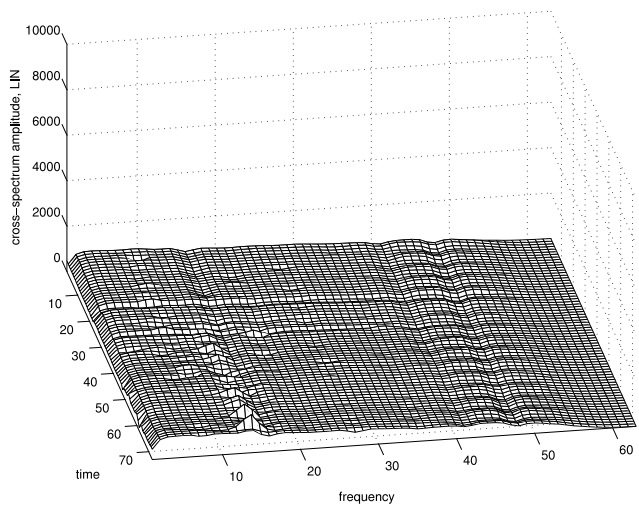

Fig. 4e

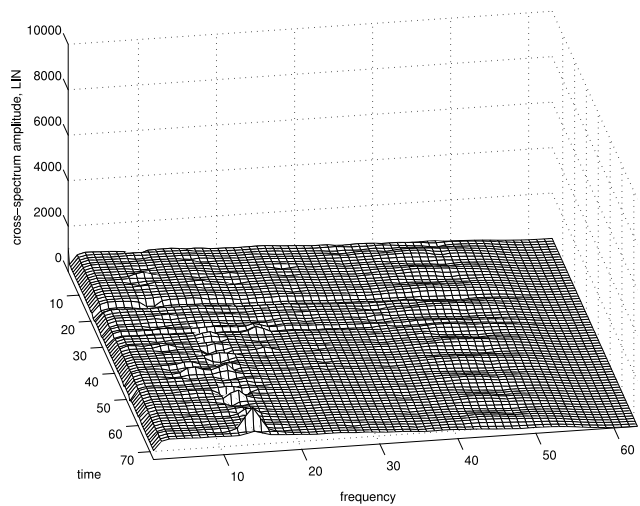

Fig. 4f

Fig. 4.—Observation at WSRT No. 10308700, 2003 December 8, source 3C286, 355 MHz, bandwidth 20 MHz, DZB correlator, 60 s average time, 73 records (≈1.2h). RFI signals were suppressed at band 2 (this figure) and not suppressed at band 3 (Fig. 5). The time-frequency presentations, *linear scale*, of the *cross-spectrum amplitudes* are given: 5X6X, 5X7X, 5X9X, 6X7X, 6X9X, and 7X9X, band 2, with RFIMS. The figures are based on the actual astronomical data.

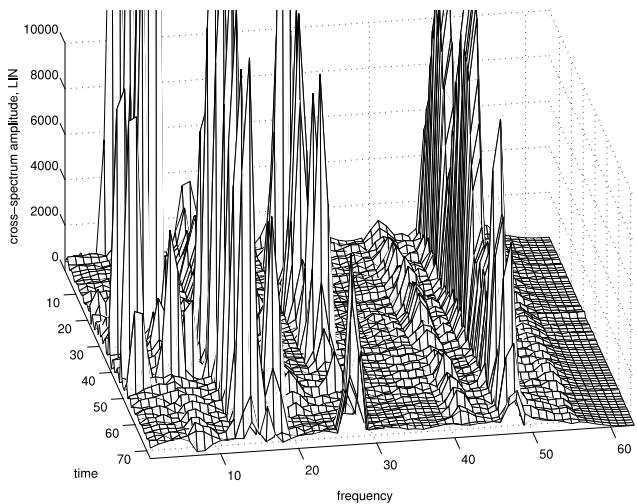
Fig. 5a

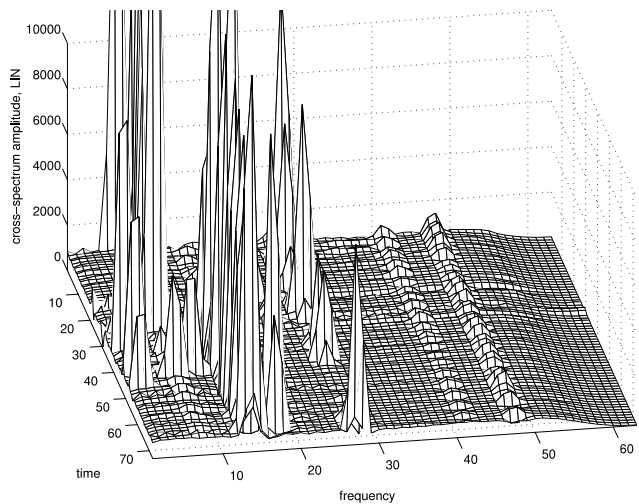
Fig. 5b

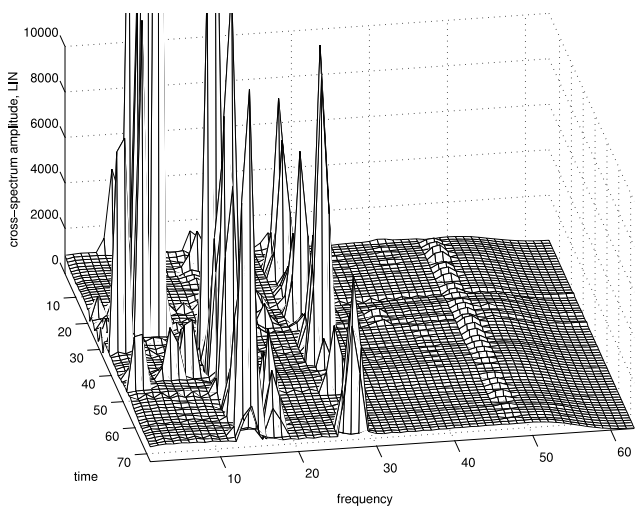
Fig. 5c

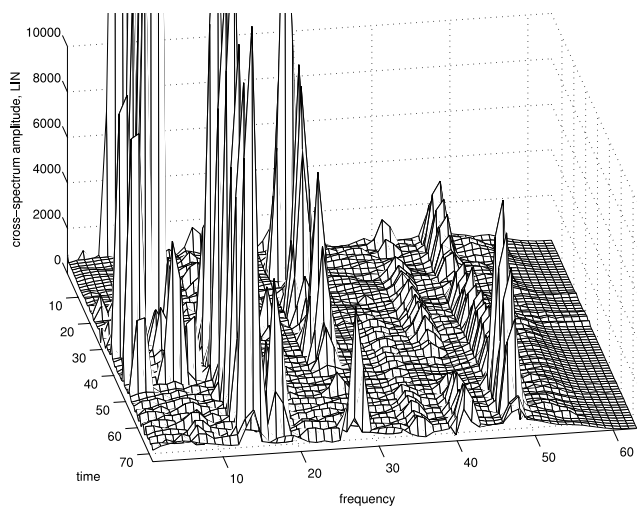
Fig. 5d

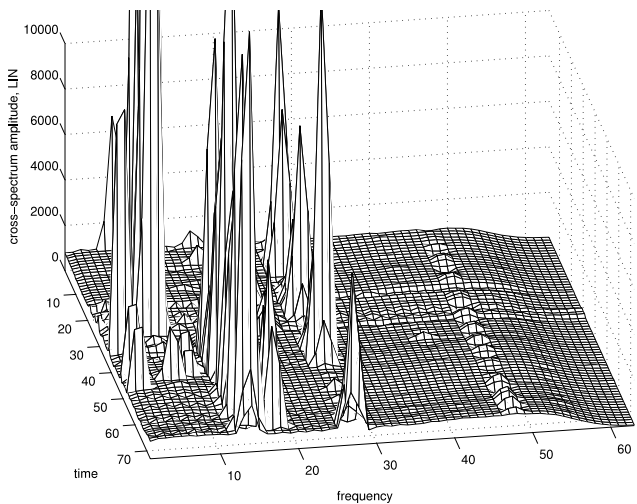
Fig. 5e

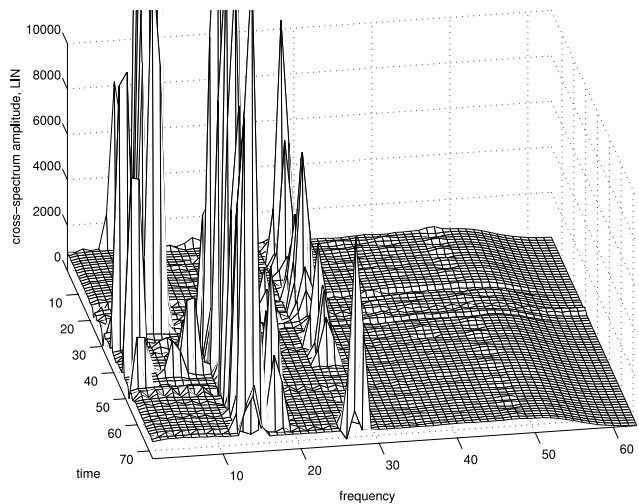
Fig. 5f

Fig. 5.—Observation at WSRT No. 10308700, 2003 December 8, source 3C286, 355 MHz, bandwidth 20 MHz, DZB correlator, 60 s average time, 73 records (≈1.2h). RFI signals were suppressed at band 2 (Fig. 4) and not suppressed at band 3 (this figure). The time-frequency presentations, *linear scale*, of the *cross-spectrum amplitudes* are given: 5X6X, 5X7X, 5X9X, 6X7X, 6X9X, and 7X9X, band 3, no RFIMS. The figures are based on the actual astronomical data.

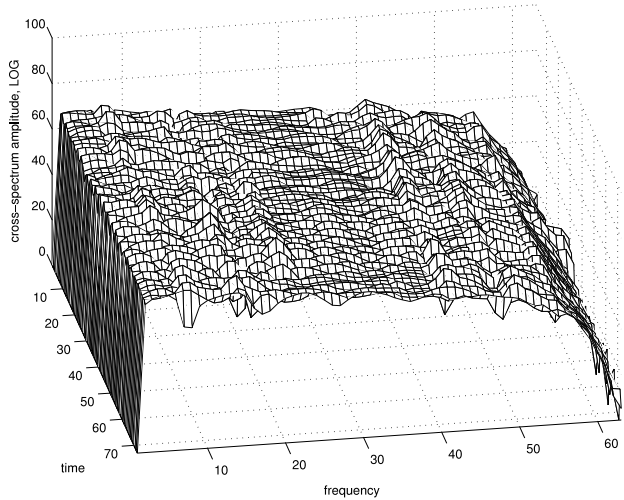

Fig. 6a

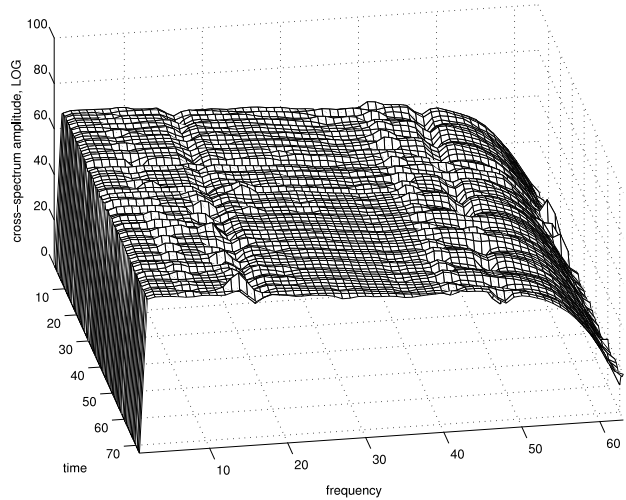

Fig. 6b

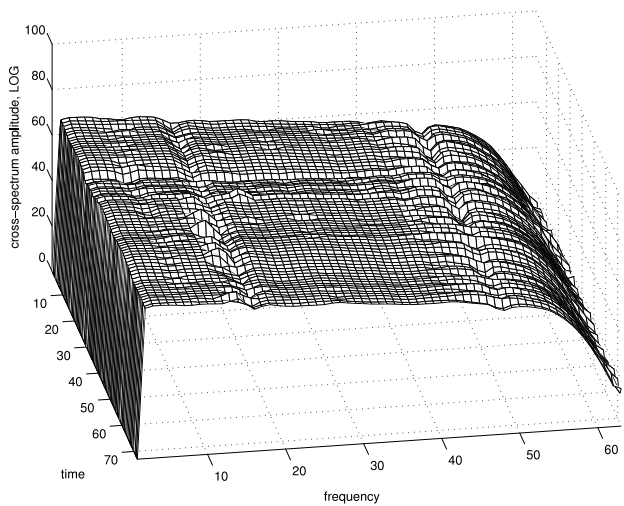

Fig. 6c

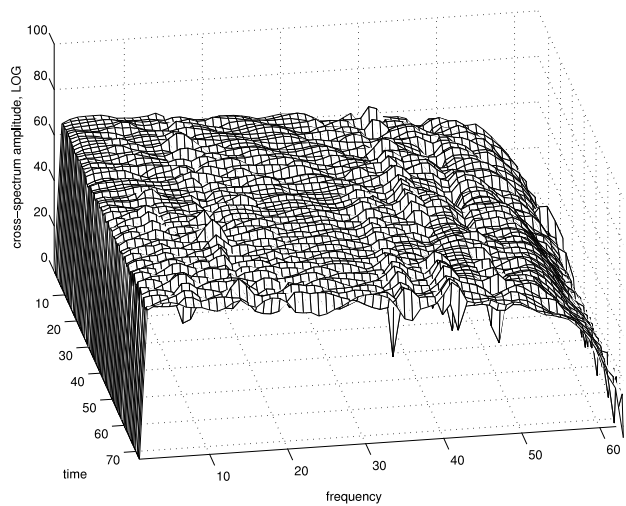

Fig. 6d

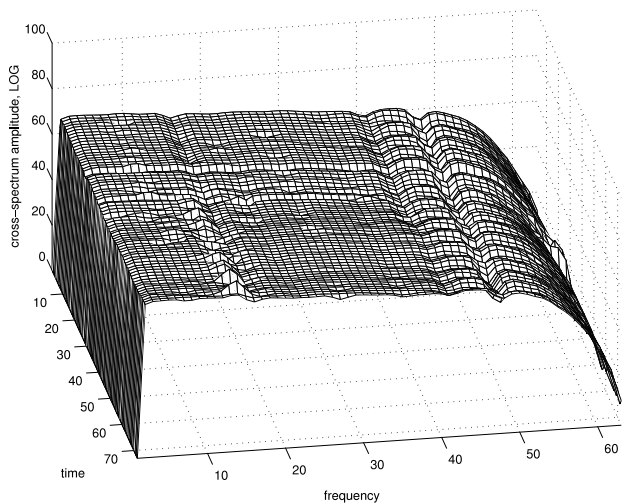

Fig. 6e

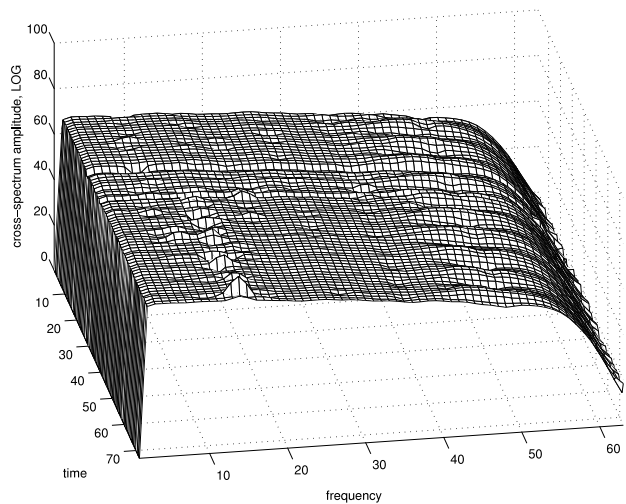

Fig. 6f

Fig. 6.—Observation at WSRT No. 10308700, 2003 December 8, source 3C286, 355 MHz, bandwidth 20 MHz, DZB correlator, 60 s average time, 73 records ($\approx 1.2h$). RFI signals were suppressed at band 2 (this figure) and not suppressed at band 3 (Fig. 7). The time-frequency presentations, *logarithmic scale*, of the *cross-spectrum amplitudes* are given: 5X6X, 5X7X, 5X9X, 6X7X, 6X9X, and 7X9X, band 2, with RFIMS. The figures are based on the actual astronomical data.

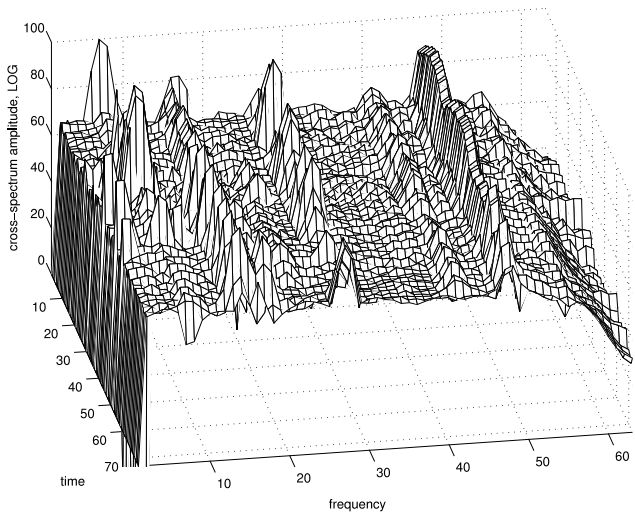

Fig. 7a

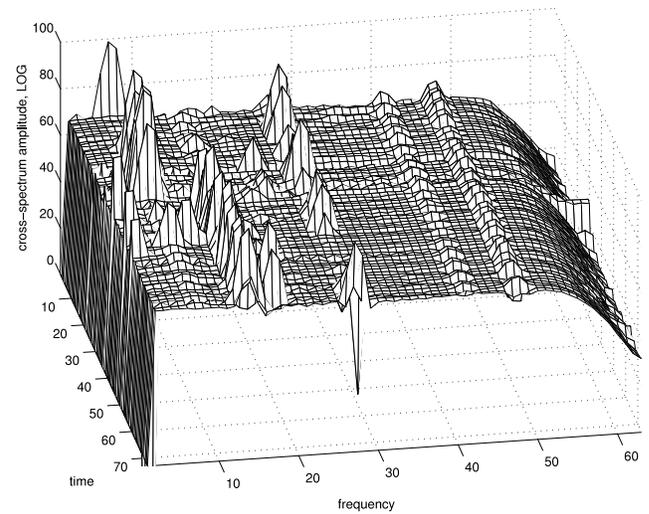

Fig. 7b

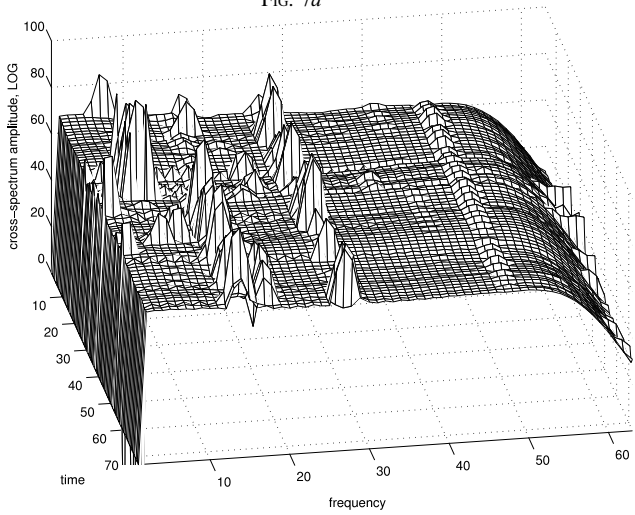

Fig. 7c

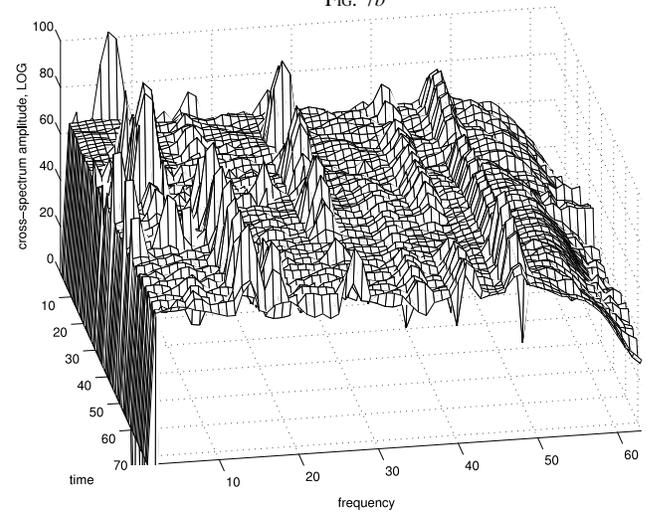

Fig. 7d

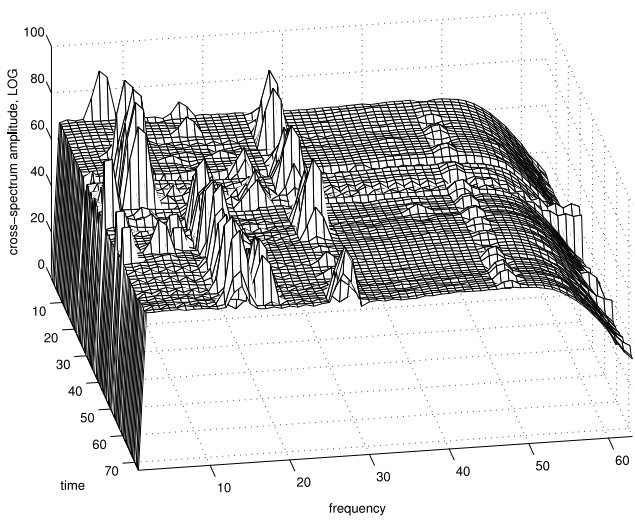

Fig. 7e

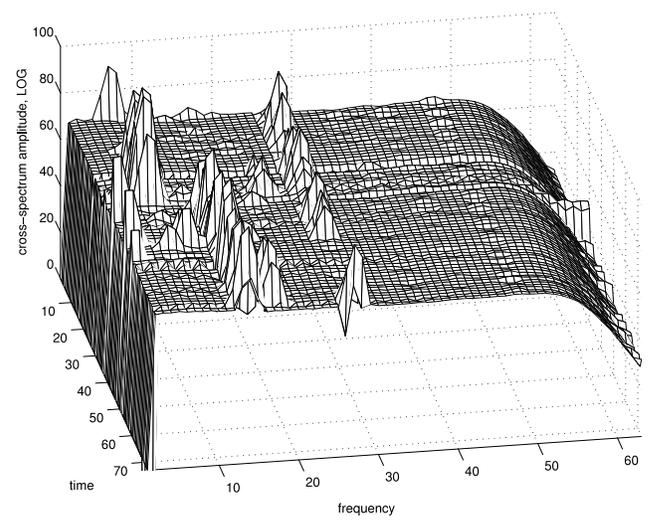

Fig. 7f

Fig. 7.—Observation at WSRT No. 10308700, 2003 December 8, source 3C286, 355 MHz, bandwidth 20 MHz, DZB correlator, 60 s average time, 73 records ($\approx$1.2h). RFI signals were suppressed at band 2 (Fig. 6) and not suppressed at band 3 (this figure). The time-frequency presentations, *logarithmic scale*, of the *cross-spectrum amplitudes* are given: 5X6X, 5X7X, 5X9X, 6X7X, 6X9X, and 7X9X, band 3, no RFIMS. The figures are based on the actual astronomical data.

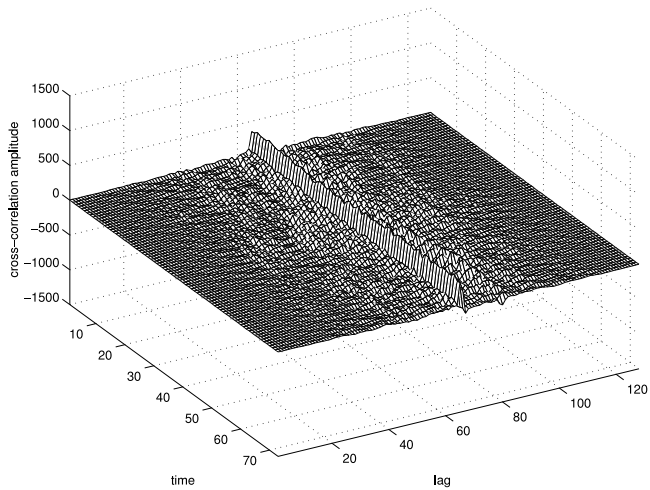

Fig. 8a

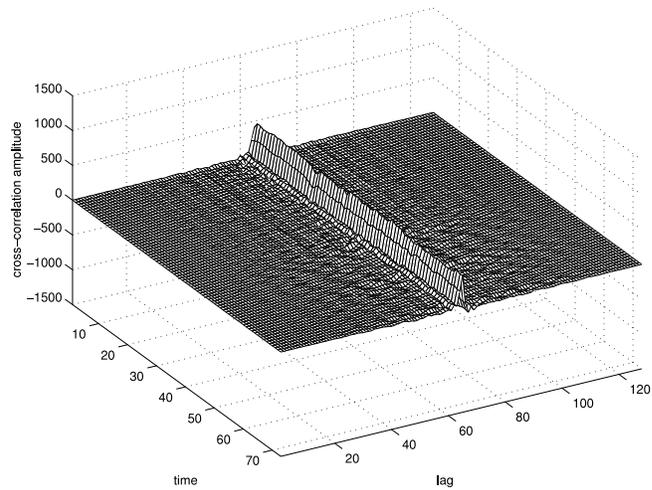

Fig. 8b

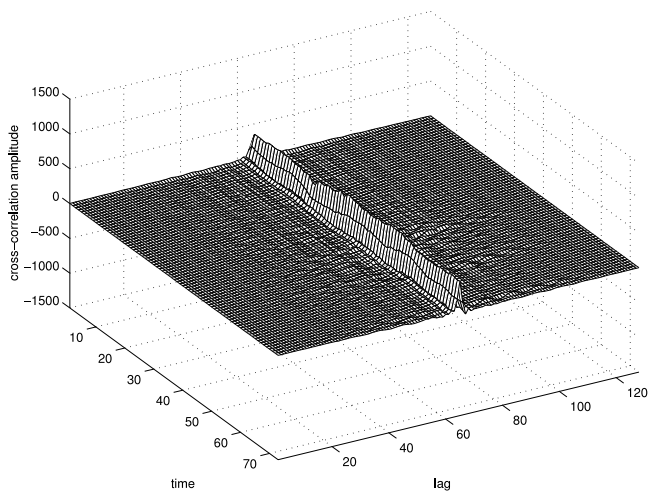

Fig. 8c

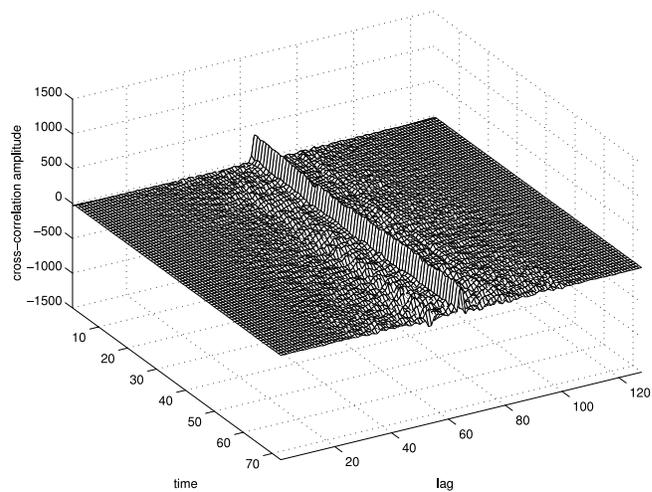

Fig. 8d

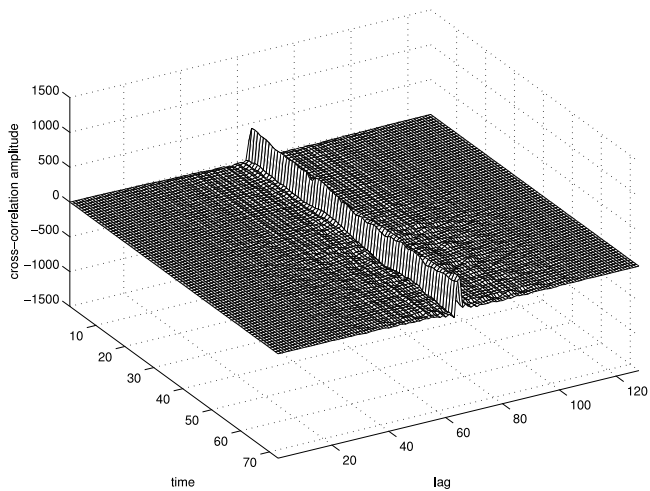

Fig. 8e

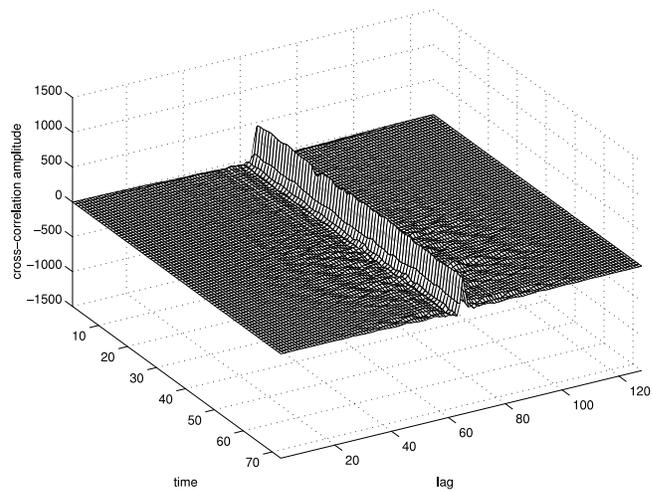

Fig. 8f

Fig. 8.—Observation at WSRT No. 10308700, 2003 December 8, source 3C286, 355 MHz, bandwidth 20 MHz, DZB correlator, 60 s average time, 73 records ($\approx 1.2h$). RFI signals were suppressed at band 2 (this figure) and not suppressed at band 3 (Fig. 9). The time-frequency presentations of the *cross-correlation functions* are given: 5X6X, 5X7X, 5X9X, 6X7X, 6X9X, and 7X9X, band 2, with RFIMS. The figures are based on the actual astronomical data.

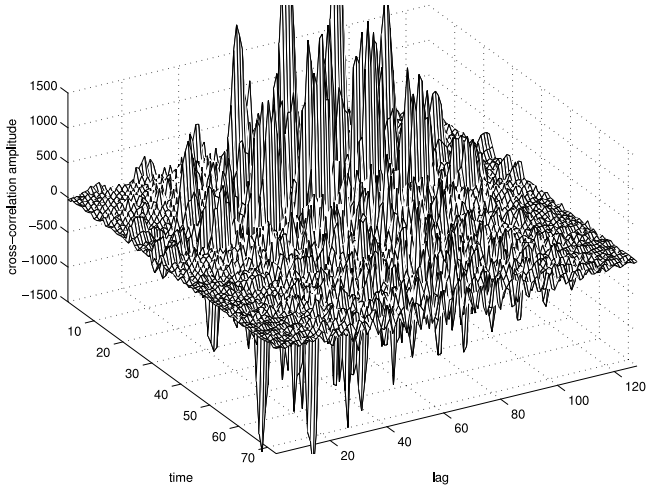

Fig. 9a

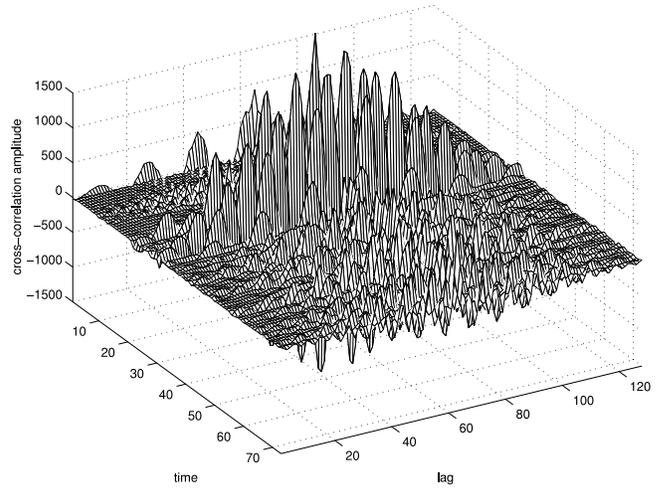

Fig. 9b

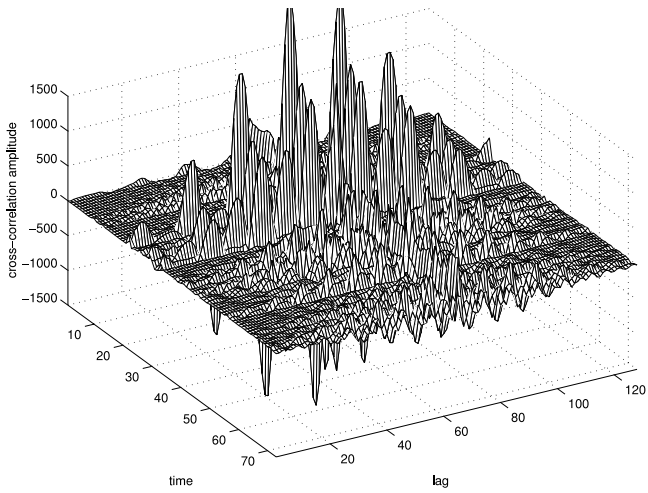

Fig. 9c

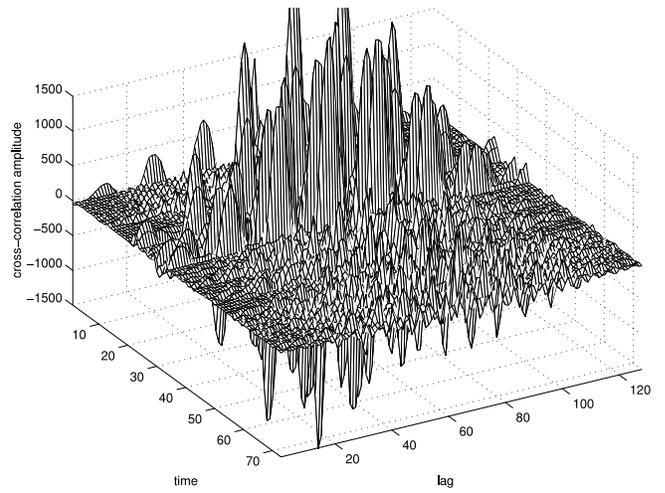

Fig. 9d

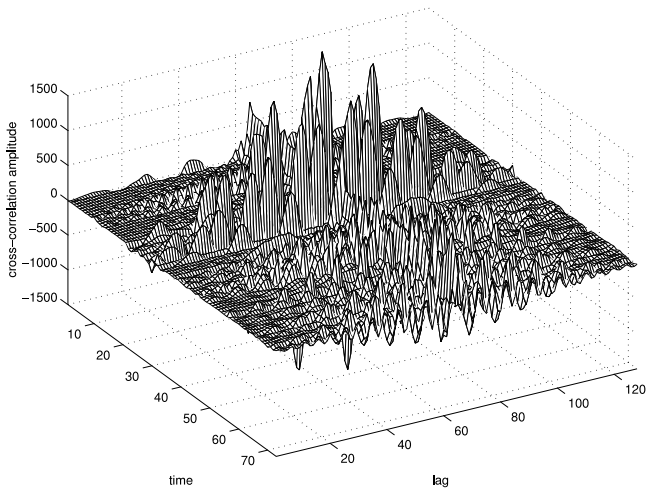

Fig. 9e

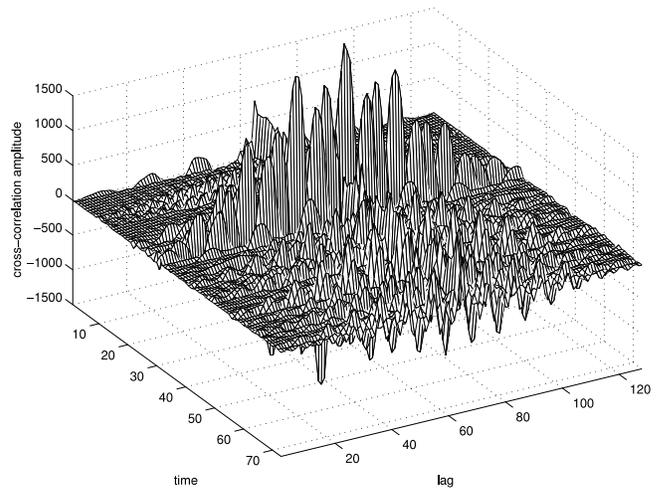

Fig. 9f

Fig. 9.—Observation at WSRT No. 10308700, 2003 December 8, source 3C286, 355 MHz, bandwidth 20 MHz, DZB correlator, 60 s average time, 73 records ($\approx 1.2h$). RFI signals were suppressed at band 2 (Fig. 8) and not suppressed at band 3 (this figure). The time-frequency presentations of the *cross-correlation functions* are given: 5X6X, 5X7X, 5X9X, 6X7X, 6X9X, and 7X9X, band 3, no RFIMS. The figures are based on the actual astronomical data.

transformed using FFT routines. The spectral signal function $sp_i(f)$ of the $i$th antenna at each discrete frequency $f$ is "cleaned" using a reference signal vector $r_n$ that is formed as a linear combination of the two "neighbors" signals from the $(i-1)$th and $(i+1)$th antennas. This procedure provides the cancellation of the useful signal from a radio source itself, such that

$$r_n(f) = [sp_i(f) - \overleftarrow{sp_{i-1}}(f) \ \overrightarrow{sp_{i+1}}(f) - sp_i(f)]^T, \quad (1)$$

$$\widehat{sp_i(f)}_n = sp_i(f)_{n-1} - w_n^T(f) r_n(f), \quad (2)$$

$$w_n(f) = w_{n-1}(f) + \mu \cdot \widehat{sp_i(f)}_{n-1} \cdot \overline{r_n(f)}. \quad (3)$$

In these formulae the functions $sp_{i-1}(f)_n, sp_i(f)_n$ and $sp_{i+1}(f)_n$ are the spectral transforms of the baseband signals $x_{i-1}(t), x_i(t)$ and $x_{i+1}(t)$ from antennas $i-1, i$ and $i+1$. The index $n$ denotes the tag number of the spectrum along the time axis or the number of iterations of the least-mean-square (LMS) algorithm depicted in equations (1), (2), and (3). The vector $r_n$ serves as the reference signal vector and the arrows above $sp_{i-1}(f)$ and $sp_{i+1}(f)$ denote the time shift (actual delay compensation) of these signals. The time shift is necessary to obtain totally coherent versions of the signal with respect to the signal of interest from the astronomical data before the actual subtraction is executed in equation (1). This operation is similar to the delay compensation operation routinely done in the correlator. The determination of the delay compensation is under computer control and the values are precalculated for each particular "source-array" geometry. Therefore, the useful signal from the radio source is not present in the vector $r_n(f)$. On the other hand, the "clean" signal $\widehat{sp_i(f)}_n$ (with overline) for each iteration $n$ is determined as an *error signal* using an LMS algorithm with a gain factor $\mu$ and the weight column vector $w_n(f)$. The overline bar in $\overline{r_n(f)}$ denotes a complex conjugation; the superscript $T$ denotes a transposition.

The number of antennas taking part in the formation of the reference channel can be larger than two, using the method described above. The vector $r_n$ may therefore contain more differences of the kind $sp_i - sp_{i-1}$. A more accurate vector $r_n$ would increase the level of RFI suppression, but the effectiveness of such steps depends on the degree of RFI correlation at the array's antennas. Thus, for radio interferometry using longer baselines, the RFI at widely separated antennas may be considerably uncorrelated.

An RFIMS pilot system for four IF channels has been completed for the WSRT, and we present some results of the test observations in Figures 3, 4, 5, 6, 7, 8, and 9.

WSRT is a 14-element correlation interferometer, and the correlator output is the primary data for further processing. Figure 3 shows the cross-correlation function, corresponding to the antenna combination RT5-RT7 with RFI mitigation and RT4-RT6 without RFI mitigation.

Figures 4, 5, 6, and 7 demonstrate cross-spectra (in linear and logarithmic scales) for RT3, RT5, RT6, and RT7, and Figures 8 and 9 give cross-correlation functions. Figures 4, 6, and 8 correspond to band 2 with RFI mitigation, while Figures 5, 7, and 9 correspond to band 3 without RFI mitigation. Both bands were tuned to the same sky frequency. RFI are sporadic and with strong variability. Nevertheless, a significant improvement is obtained, which is useful for continuum observations.

## 4. CONCLUSIONS

First results obtained with the real-time RFI mitigation system based on FPGAs are encouraging. Application of modern FPGAs allows implementing signal-processing algorithms with sufficient complexity to get significant RFI mitigation effects.

The FPGAs are reprogrammable devices, and this feature may be used to implement different algorithms, depending on the nature of RFI and on the type of observation being done.

## APPENDIX A

## GAIN AND LOSS AFTER RFI EXCISION

Let us first consider the "ideal" situation: all RFI are 100% detected and deleted. What are the gain and loss benefits? Let $\mathcal{P}_{sig}$, $\mathcal{P}_{sys}$, $\mathcal{P}_{RFI}$ be antenna signal, system noise and RFI spectral power densities, respectively, $\alpha = \mathcal{P}_{RFI}/\mathcal{P}_{sys}$. The time-frequency plane ($t \otimes f$-plane, Fig. 10) represents the power spectrum after the short-time Fourier transform (STFT). STFT is made with the numbers obtained after sampling with a rate $f_s$, and the length of the transform is $N_{FT}$; each frequency point corresponds to the time interval $t_F = N_{FT}(1/f_s)$ (time resolution). The total number of points on the $t \otimes f$-plane is $M$. The fractional RFI area is $\beta = S_{RFI}/S_{tot}$, $S_{RFI}$ is the total area occupied by RFI, and $S_{tot}$ is the total square of the part of the $t \otimes f$-plane that is used for the integration over $M$ points (to get a gain $\sim \sqrt{M}$ in a signal-to-noise ratio [S/N] after correlation or total power detection). The value $\beta$ is less than 1, $0 \leq \beta < 1$, because if $\beta = 1$, there is no reason to apply excision (the signal will be totally excised as well).

The rms at the output of correlator (or TPD) in the absence of RFI and $\mathcal{P}_{sig} \ll \mathcal{P}_{sys}$ is $rms_0 = \mathcal{P}_{sys}/\sqrt{M}$, and the S/N in the absence of RFI is $S/N_0 = \mathcal{P}_{sig}/rms_0 = (\mathcal{P}_{sig}/\mathcal{P}_{sys}) \sqrt{M}$.

The rms in the presence of RFI is $rms_{RFI} = [(\mathcal{P}_{sys}/\sqrt{M})^2 + \beta(\mathcal{P}_{RFI})^2]^{1/2} = P_{sys}[(1/M) + \beta\alpha^2]^{1/2}$. The S/N in the presence of RFI is $S/N_{RFI} = (\mathcal{P}_{sig}/\mathcal{P}_{sys})\{1/[(1/M) + \beta\alpha^2]^{1/2}\}$. We suppose here that the RFI is not reduced after integration (100% correlated, the worst case).

The rms after RFI excision is $rms_{EXC} = \mathcal{P}_{sys}/[M(1 - \beta)]^{1/2}$, and the S/N after RFI excision is $S/N_{EXC} = (\mathcal{P}_{sig}/\mathcal{P}_{sys})[M(1 - \beta)]^{1/2}$.

Now the gain (due to RFI suppression) is $G_{EXC} = (S/N_{EXC})/(S/N_{RFI}) = (1/M + \beta\alpha^2)^{1/2}[M(1 - \beta)]^{1/2}$, and the loss is $R_{EXC} = (S/N)_{EXC}/(S/N)_0 = (1 - \beta)^{1/2}$.

Figure 11 shows how $G_{EXC}$ depends on $\beta$ (fraction of RFI on the $t \otimes f$-plane or duty cycle), while $\alpha$ (RFI intensity relative to the system noise intensity) is a free parameter. The loss $R_{EXC}$ is also represented in this figure (*dashed line*).

Until now it was assumed that RFI is 100% correlated at the different sites of a radio interferometer and the correlator output contains a harmful bias due to RFI (dc component). This is also valid for *single-dish* observations. For a totally uncorrelated RFI (large interferometers, VLBI) $rms_{RFI} = \{[P_{sys}^2 + \beta(P_{RFI})^2]^{1/2}\}(1/\sqrt{M})$ and $G_{EXC} = (1 + \beta\alpha^2)^{1/2}(1 - \beta)^{1/2}$.

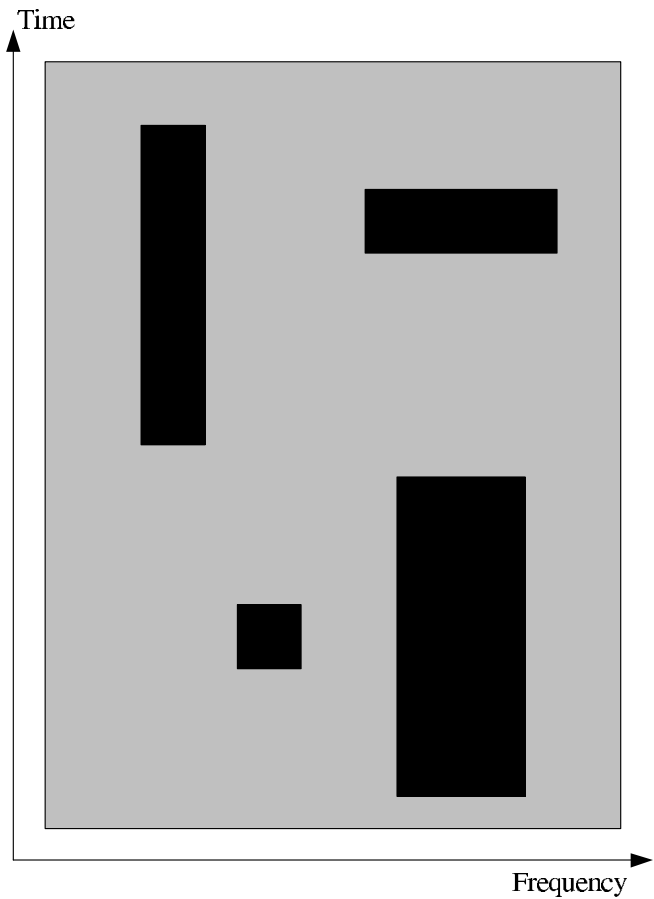

Fig. 10.—Time-frequency plane: system noise (*gray*); RFI (*black*).

The formulae above presume that the parts of data with RFI, $\beta M$ samples are totally deleted. But sometimes this operation is not possible because of technical constraints in the existing hardware. So there is another possibility: to substitute these $\beta M$ samples by the noise-like numbers, for which the variance is approximately the same as for the pure $P_{\text{sys}}$ (substitution by zeros is bad because of the undesirable bias in the correlator output). In this case, the rms after RFI excision is $\text{rms}_{\text{EXC}} = P_{\text{sys}}/\sqrt{M}$, the S/N after the RFI excision is $\text{S/N}_{\text{EXC}} = (P_{\text{sig}}/P_{\text{sys}})(1-\beta)\sqrt{M}$, and the signal of interest is reduced by a factor $(1-\beta)$.

## APPENDIX B

## DELAY IN RFI DETECTION

Now let us go to the more realistic situation involving nonideal RFI detection (and hence excision). There is always a delay before RFI is detected, which could be considerable for weak RFI. Let the probability distribution of the system noise on the $t \otimes f$-plane in the absence of RFI be Gaussian with the mean $\mu_0 = 0$ and the standard deviation $\sigma = 1$. Let $h$ be the one-side normalized detection threshold, so the probability of false alarm $p_{\text{FA}} = 1 - \Phi(h, 0, 1)$, where $\Phi(x, \mu, \sigma)$ is the cumulative Gaussian distribution with mean $\mu$ and variance $\sigma^2$. A detection delay or an averaged run length (ARL) in the absence of RFI is $\text{ARL} = 1/p_{\text{FA}}$. In the presence of RFI a bias $\mu_{\text{RFI}}$ appears and the ARL in this case is $\text{ARL} = 1/[1 - \Phi(h, \mu_{\text{RFI}}, 1)]$, $\mu_{\text{RFI}} = \mathcal{P}_{\text{RFI}}/\sigma$. Figure 12 shows how the detection probability $p_{\text{det}}$ and ARL depend on $\mu_{\text{RFI}}$ for different thresholds, $h = 1.7(p_{\text{FA}} = 0.046)$, $h = 2.4(p_{\text{FA}} = 8.42 \times 10^{-3})$, $h = 3(p_{\text{FA}} = 1.4 \times 10^{-3})$. The ARL in the presence of RFI corresponds to the number of RFI samples "unnoticed" by the detection procedure. All samples that are larger than the threshold are deleted, but those in the ARL regions remain. Therefore, the rms after RFI excision is $\text{rms}_{\text{EXC}} = (\{\mathcal{P}_{\text{sys}}/[M(1-\beta p_{\text{det}})]^{1/2}\}^2 + \beta(1-p_{\text{det}})(\alpha \mathcal{P})^2 \text{sys})^{1/2}$, $p_{\text{det}} = 1 - \Phi(h, \mu_{\text{RFI}}, 1)$. Figure 13 shows the gain $G_{\text{EXC}}$ and loss $R_{\text{EXC}}$ versus $\alpha$ in this case for $\beta = 0.3$, $h = 2(P_{\text{FA}} = 0.046)$. The dashed line shows the gain when RFI detection is made after coherently adding all equal RFIs from 14 antennas. In this case $\text{ARL}_{14} = 1/[1 - \Phi(h, \mu_{\text{RFI}}\sqrt{14}, 1)]$.

## APPENDIX C

## ANALYSIS OF THE ADAPTIVE INTERFERENCE CANCELLER

A simplified analysis of the interference canceller with a reference channel is given below; see the setup in Figure 14. There are two inputs: a signal from a radio source $S$ and an interference signal $R$. These two signals, uncorrelated with each other, are mixed together through the system functions $H(z)$ and $G(z)$ and the products appear in the *main input d* and the *reference input x*. The transfer functions $H(z)$ and $G(z)$ depict the difference of levels of interference in these two channels [$G(z)$] and the leakage or penetration of the useful signal $S$ in the reference channel [$H(z)$]. There are also additive noise signals (system noise) $N$ and $M$ added to $S$ and $R$, respectively. The reference input is transformed in an optimal filter $W(z)$ (Wiener filter), so that the difference between the main input and the filter output would be minimal in the mean-squared sense. The output of the subtractor is the interference canceller output. The value of the signal-to-"noise+interference" at the subtractor output is estimated and compared with the S/N at the input of noise canceller both in the presence and the absence of interference ("ideal" case).

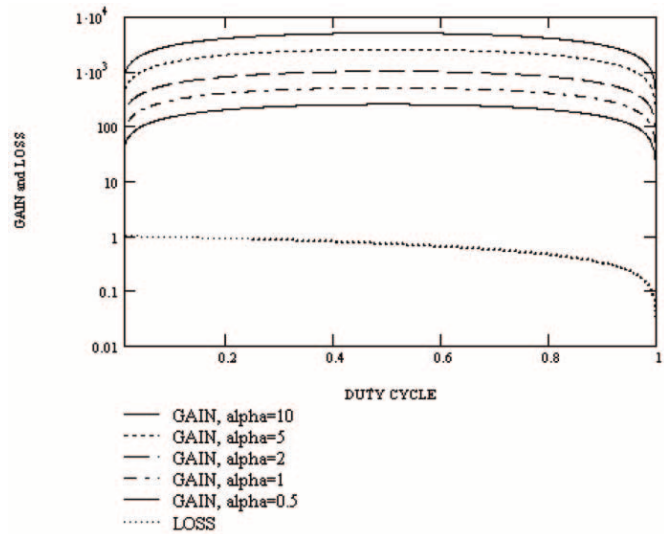

Fig. 11.—Gain and loss vs. RFI duty cycle $\beta$: fractional RFI intensity $\alpha$ is a free parameter; RFIs are 100% correlated at the radio interferometer sites; integration factor $M = 10^6$.

The main signal input is

$$d(t) = S(t) + N(t) + g(t) \times R(t), \tag{C1}$$

and the reference signal input is

$$x(t) = R(t) + M(t) + h(t) \times S(t). \tag{C2}$$

Using "spectral" language allows us to estimate interference canceller behavior in the frequency domain, which is well suited to the principal application area of such a canceller—spectral observations with a persistent (in time) interference. The spectral densities (C1) and (C2) may be rewritten as

$$\Phi_d(z) = \Phi_S(z) + \Phi_N(z) + \Phi_R(z)|G(z)|^2, \tag{C3}$$

$$\Phi_x(z) = \Phi_R(z) + \Phi_M(z) + \Phi_S(z)|H(z)|^2. \tag{C4}$$

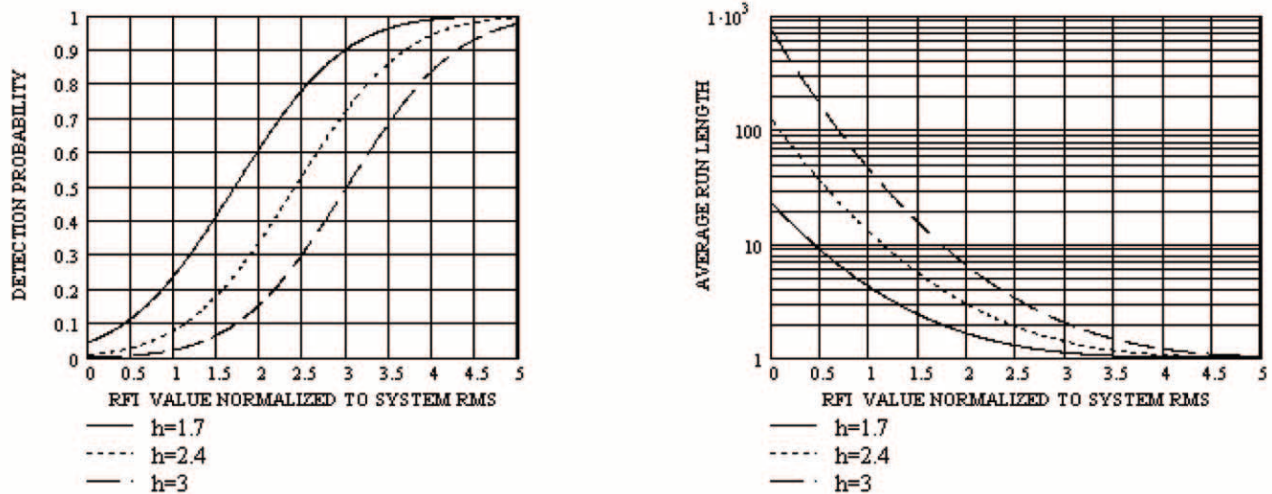

Fig. 12.—Probability detection (left) and average run length (right) vs. RFI fractional intensity. Gaussian distribution for the system noise; threshold $h$ is the free parameter.

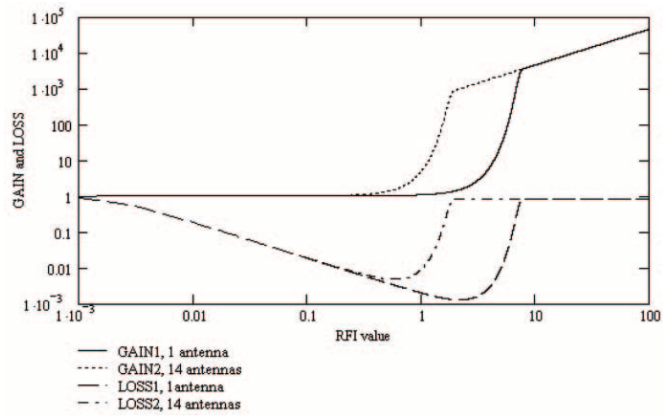

Fig. 13.—Gain and loss after the RFI excision vs. RFI fractional intensity, Gaussian distribution for the system noise; threshold $h = 2$; false alarm probability $P_{fa} = 0.046$; RFI duty cycle $\beta = 0.3$; integration factor $M = 10^6$; RFIs are 100% correlated.

The Wiener filter transfer function for the interference estimation is

$$W(z) = \frac{\Phi_{dx}(z)}{\Phi_x(z)} = \frac{\overline{H(z)}\Phi_S(z) + G(z)\Phi_R(z)}{\Phi_R(z) + \Phi_M(z) + \Phi_S(z)|H(z)|^2}, \qquad (C5)$$

where $\Phi_{dx}(z)$ is the cross-spectrum of $d$ and $x$. The effectiveness of the interference canceller depends on the transfer functions $H$ and $G$. The output $e$ of the subtractor consists of the two parts: one is determined by the signal of interest

$$\Phi_{e.S}(z) = |1 - H(z)W(z)|^2 \Phi_S(z), \qquad (C6)$$

and another is determined by the interference

$$\Phi_{e.R}(z) = |G(z) - W(z)|^2 \Phi_R(z) + |W(z)|^2 \Phi_M(z). \qquad (C7)$$

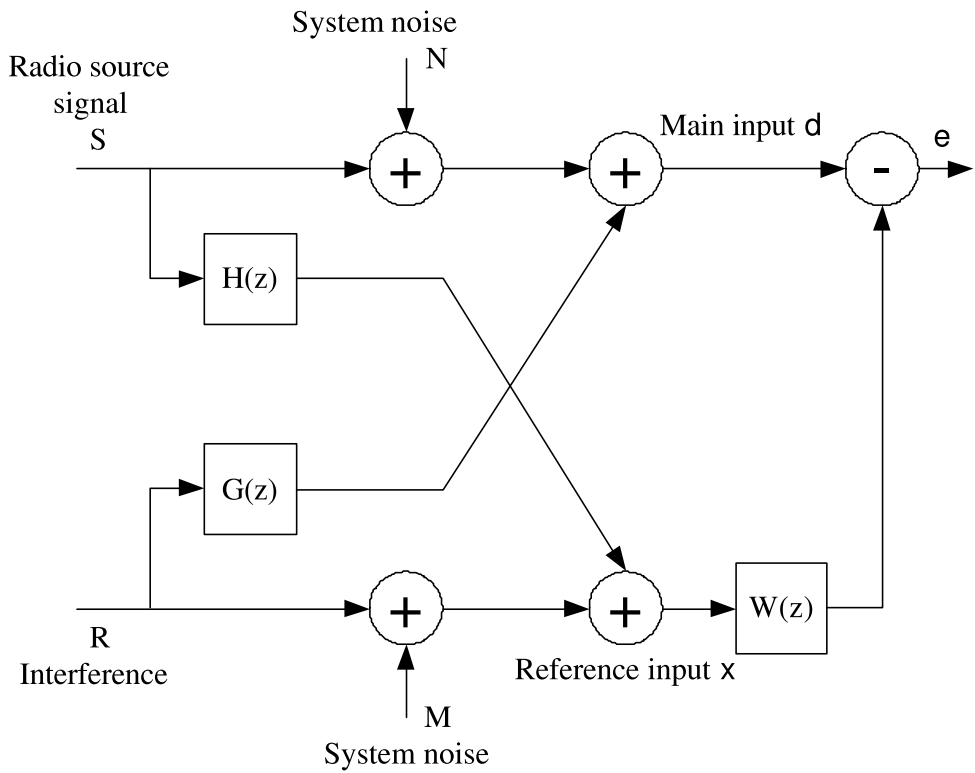

Fig. 14.—RFI canceller setup.

Replacing $W(z)$ in equations (C6) and (C7) by equation (C5) we get

$$\Phi_{e.S}(z) = \Phi_S(z)\left|1 - H(z)\frac{\overline{H(z)}\Phi_S(z) + G(z)\Phi_R(z)}{\Phi_R(z) + \Phi_M(z) + \Phi_S(z)|H(z)|^2}\right|^2, \tag{C8}$$

$$\Phi_{e.R}(z) = \Phi_R(z)\left|G(z) - \frac{\overline{H(z)}\Phi_S(z) + G(z)\Phi_R(z)}{\Phi_R(z) + \Phi_M(z) + \Phi_S(z)|H(z)|^2}\right|^2 + \Phi_M(z)\left|\frac{\overline{H(z)}\Phi_S(z) + G(z)\Phi_R(z)}{\Phi_R(z) + \Phi_M(z) + \Phi_S(z)|H(z)|^2}\right|^2. \tag{C9}$$

Now we define the S/N for the spectral densities at the main input, reference input, and canceller output, respectively:

$$\text{S/N}_{\text{main}}(z) = \frac{\Phi_S(z)}{|G(z)|^2\Phi_R(z) + \Phi_N(z)}, \tag{C10}$$

$$\text{S/N}_{\text{ref}}(z) = \frac{\Phi_S(z)|H(z)|^2}{\Phi_R(z) + \Phi_M(z)}, \tag{C11}$$

$$\text{S/N}_{\text{out}}(z) = \frac{\Phi_{e.S}(z)}{\Phi_{e.R}(z) + \Phi_N(z)}. \tag{C12}$$

Now let us consider several examples.

1. In the textbooks (Haykin 1996; Widrow & Stearns 1985) the additive noises are often absent, $\Phi_N(z) = \Phi_M(z) = 0$; that is, it is assumed that the signal of interest $S$ and interference $R$ are both much stronger than system noise $N$ and $M$, respectively. In this case

$$\text{S/N}_{\text{out}}(z) = \frac{\Phi_S(z)\left|1 - H(z)\frac{\overline{H(z)}\Phi_S(z) + G(z)\Phi_R(z)}{\Phi_R(z) + \Phi_S(z)|H(z)|^2}\right|^2}{\Phi_R(z)\left|G(z) - \frac{\overline{H(z)}\Phi_S(z) + G(z)\Phi_R(z)}{\Phi_R(z) + \Phi_S(z)|H(z)|^2}\right|^2} = \frac{\Phi_R(z)}{\Phi_S(z)|H(z)|^2} = \frac{1}{\text{S/N}_{\text{ref}}(z)}, \tag{C13}$$

which is known as a *power inversion*: the weaker the signal of interest, the better the interference suppression.

2. Let $H(z) = 0$ (no signal leakage in the reference channel); that is, we have a separate "off-source" antenna for interference monitoring and $\Phi_M(z) > 0$:

$$\text{S/N}_{\text{out}}(z) = \frac{\Phi_S(z)[\Phi_R(z) + \Phi_M(z)]}{G(z)^2\Phi_R(z)\Phi_M(z)}, \tag{C14}$$

which for $G(z) = 1$ is

$$\text{S/N}_{\text{out}}(z) = \frac{\Phi_S(z)[\Phi_R(z) + \Phi_M(z)]}{\Phi_R(z)\Phi_M(z)} = \frac{\Phi_S(z)}{\Phi_R(z)}[\text{INR}(z) + 1], \tag{C15}$$

where $\text{INR}(z) = \Phi_R(z)/\Phi_M(z)$.

3. It is more realistic to take into account system noise $\Phi_N(z)$ in the main channel. For $G(z) = 1$, and without interference cancellation

$$\text{S/N}_{\text{main}} = \frac{\Phi_S(z)}{\Phi_R(z) + \Phi_N(z)}. \tag{C16}$$

For $H(z) = 0$, we get the benefit of interference suppression ("gain")

$$\text{GAIN}(z) = \frac{\text{S/N}_{\text{out}}(z)}{\text{S/N}_{\text{main}}(z)} = \frac{\dfrac{\Phi_S(z)}{\Phi_R(z)\left|G(z) - \dfrac{G(z)\Phi_R(z)}{\Phi_R(z) + \Phi_M(z)}\right|^2 + \Phi_M(z)\left|\dfrac{G(z)\Phi_R(z)}{\Phi_R(z) + \Phi_M(z)}\right|^2 + \Phi_N(z)}}{\dfrac{\Phi_S(z)}{|G(z)|^2\Phi_R(z) + \Phi_N(z)}}. \tag{C17}$$

If, for simplicity, we assume that the noises in the main and reference channels have the same intensity, $\Phi_N(z) = \Phi_M(z)$ and $G(z) = 1$, we get

$$\text{GAIN}(z) = \frac{[\text{INR}(z) + 1]^2}{2\text{INR}(z) + 1}. \tag{C18}$$

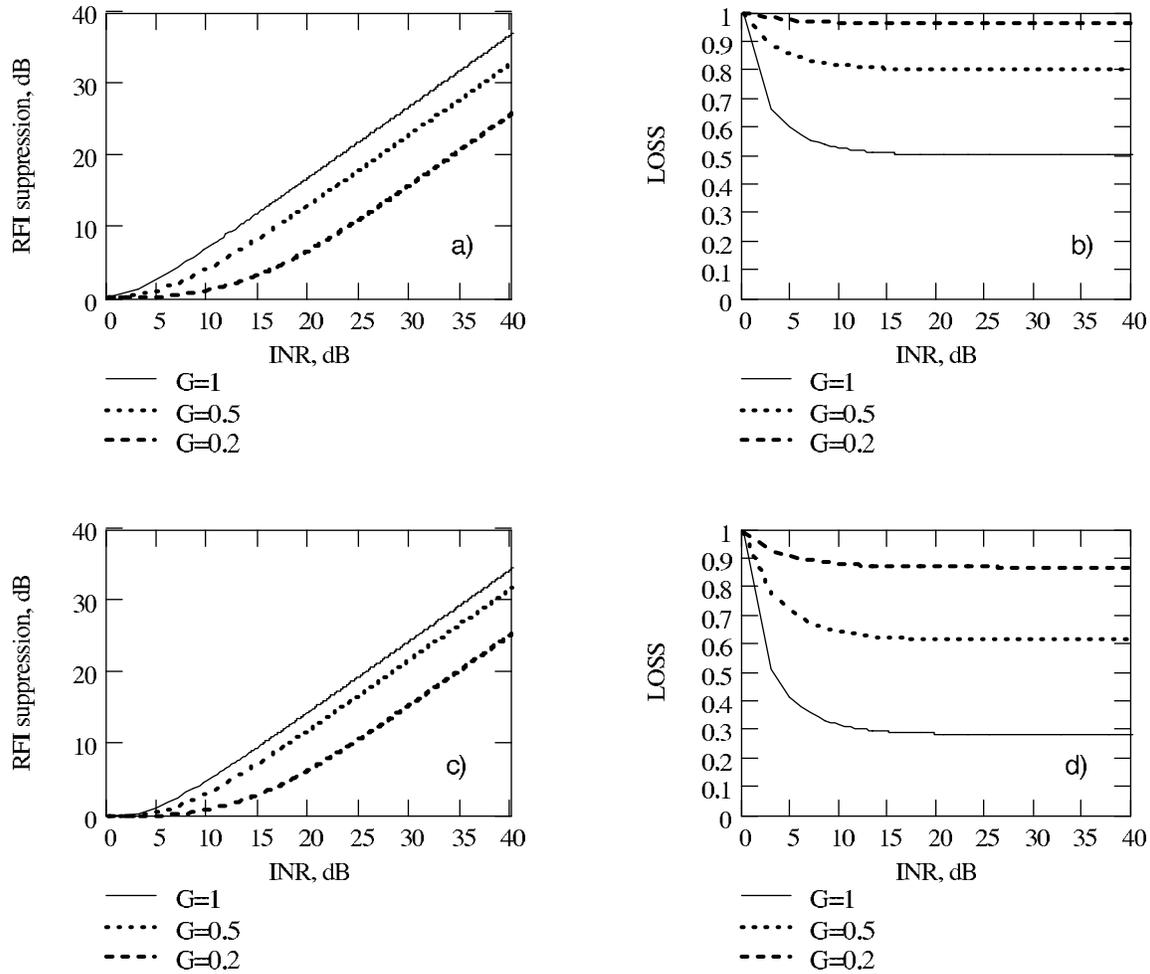

Fig. 15.—Gain and loss as a function of interference-to-noise ratio. (*a*)–(*b*) No signal leakage in the reference channel, $H = 0.0$; (*c*)–(*d*) the same, but with a certain signal leakage in the reference channel, $H = 0.25$.

Figures 15*a* and 15*b* show GAIN and $\text{LOSS}(z) = S/N_{\text{out}}(z)/[\Phi_S(z)]/[\Phi_N(z)]$ for different $G$ as a function of INR and $H = 0$. When $H > 0$, that is, some signal-of-interest leakage exists in the reference channel, a worsening of interference suppression occurs, which is illustrated in Figures 15*c* and 15*d* for $H = 0.25$. Nevertheless, even in this case a significant improvement may be achieved compared with the observations without interference cancellation.

## APPENDIX D

## RFI SUPPRESSION DUE TO FRINGE-DELAY TRACKING

A particular geometry "radio source–baseline–RFI source" may result in different RFI suppression. Here we give several examples for WSRT. The calculations are based on (Thompson 1982).

1. Narrowband RFI suppression due to modulation produced by the *fringe stopping* procedure. The RFI signal from a stationary source at the correlator multiplier output is modulated with the fringe frequency

$$F = 7.272 \times 10^{-5} D_\lambda \cos(\delta) \cos(H), \tag{D19}$$

where $D_\lambda$ is the baseline in wavelength (WSRT has an east-west orientation), $\delta$ is the radio source declination, $H$ is its hour angle. RFI attenuation after integration during time $T$ is

$$AT = \sin(\pi FT)/(\pi FT). \tag{D20}$$

Additional integration is made during an image formation due to *gridding by convolution* procedure

$$T_{\text{grid}} = \frac{\Delta u |\csc(\delta)|}{7.272 \times 10^{-5} D_\lambda (|\sin(H)| + |\cos(H)||\csc(\delta)|)}, \tag{D21}$$

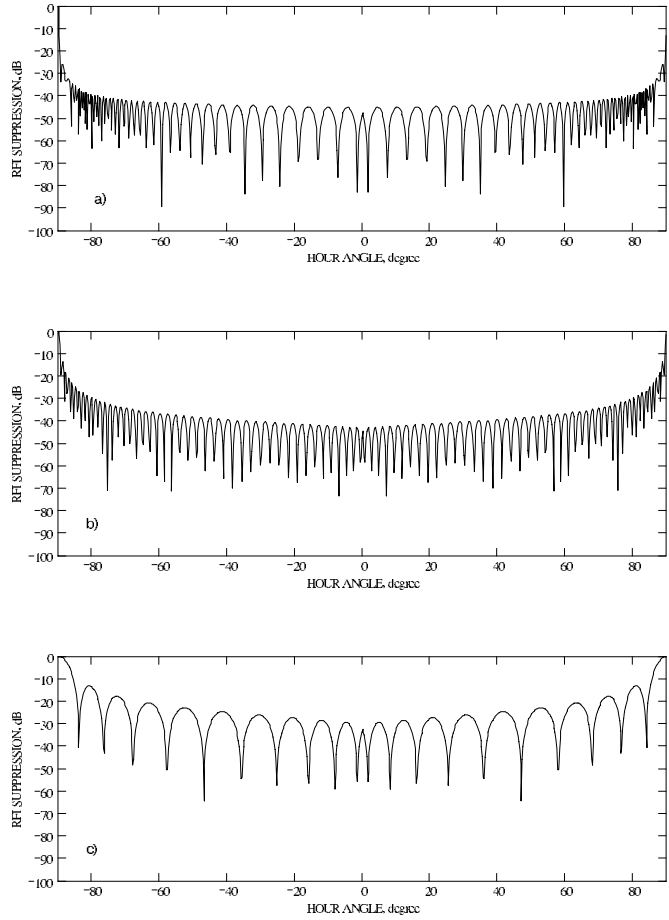

Fig. 16.—RFI suppression due to fringe stopping as a function of hour angle; (*a*) $\delta = 10°$; (*b*) $\delta = 40°$; (*c*) $\delta = 80°$; baseline $D = 2.7$ km; $d = 25$ m; $f_0 = 1420$ MHz; $\Delta u = d/\lambda = 118$.

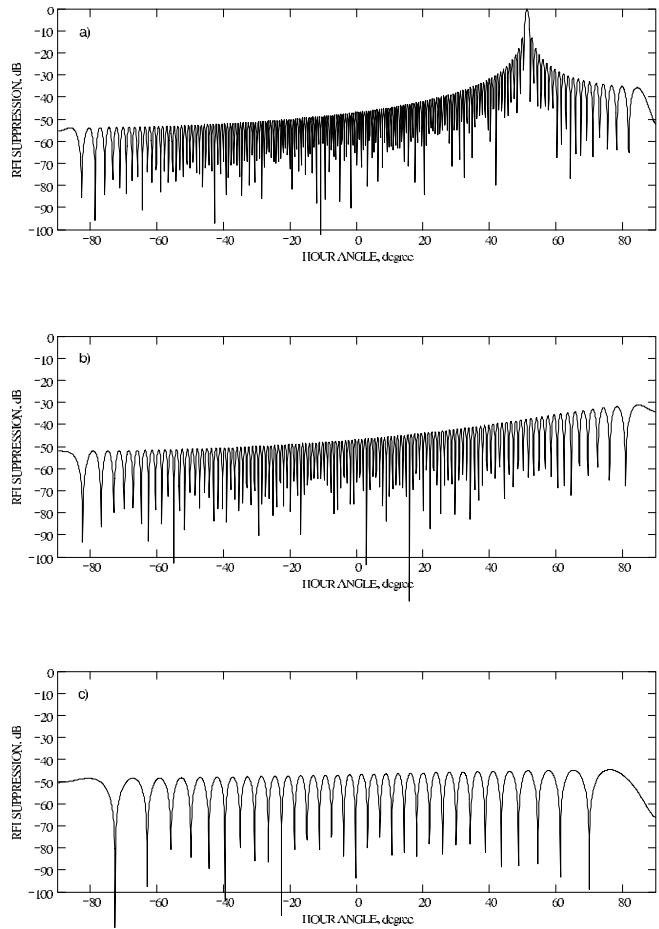

Fig. 17.—RFI suppression due to delay tracking as a function of hour angle. (a) $\delta = 10°$; (b) $\delta = 50°$; (c) $\delta = 80°$; baseline $D = 2.7$ km, RFI bandwidth $\Delta f = 10$ MHz, $\theta_{\mathrm{RFI}} = 50°$.

$\Delta u = d/\lambda$, $d$ is diameter of antenna, $d = 25$ m. Therefore, the total RFI attenuation due to fringe stopping is

$$\mathrm{AT}_{\mathrm{FR}} = \sin{(\pi \mathrm{FT}_{\mathrm{grid}}/2)}/(\pi \mathrm{FT}_{\mathrm{grid}}/2). \tag{D22}$$

2. Wide-band RFI attenuation due to the *delay tracking* procedure is

$$\mathrm{AT}_{\mathrm{delay}} = \sin{(\pi \Delta f \Delta \tau)}/(\pi \Delta f \Delta \tau), \tag{D23}$$

where

$$\Delta \tau = \tau_{\mathrm{geom}}(\mathrm{source}) - \tau_{\mathrm{geom}}(\mathrm{RFI}) = [\cos{(\delta)}\sin{(H)} - \sin{(\theta_{\mathrm{RFI}})}]D/c, \tag{D24}$$

$\theta_{\mathrm{RFI}}$ is the RFI angle, $c$ is the speed of light, and $\Delta f$ is the RFI bandwidth.

Figure 16 shows RFI attenuation due to fringe stopping for different declinations and maximum baseline $D = 2.7$ km, $f_0 = c/\lambda = 1420$ MHz.

Figure 17 shows RFI attenuation due to delay tracking for different declination and fixed position of RFI source, $\Delta f = 10$ MHz, $D = 2.7$ km.